# Ultrafast photonic reinforcement learning based on laser chaos

**Makoto Naruse[1], Yuta Terashima[2], Atsushi Uchida[2] & Song-Ju Kim[3]**

[1] *Strategic Planning Department, National Institute of Information and Communications Technology, 4-2-1 Nukui-kita, Koganei, Tokyo 184-8795, Japan*

[2] *Department of Information and Computer Sciences, Saitama University, 255 Shimo-Okubo, Sakura-ku, Saitama city, Saitama 338-8570, Japan*

[3] *WPI Center for Materials Nanoarchitectonics, National Institute for Materials Science, 1-1 Namiki, Tsukuba, Ibaraki 305-0044, Japan*

\* Corresponding author. Email: naruse@nict.go.jp

**ABSTRACT**

Reinforcement learning involves decision making in dynamic and uncertain environments, and constitutes one important element of artificial intelligence (AI). In this paper, we experimentally demonstrate that the ultrafast chaotic oscillatory dynamics of lasers efficiently solve the multi-armed bandit problem (MAB), which requires decision making concerning a class of difficult trade-offs called the exploration−exploitation dilemma. To solve the MAB, a certain degree of randomness is required for exploration purposes. However, pseudo-random numbers generated using conventional electronic circuitry encounter severe limitations in terms of their data rate and the quality of randomness due to their algorithmic foundations. We generate laser chaos signals using a semiconductor laser sampled at a maximum rate of 100 GSample/s, and combine it with a simple decision-making principle called tug-of-war with a variable threshold, to ensure ultrafast, adaptive and accurate decision making at a maximum adaptation speed of 1 GHz. We found that decision-making performance was maximized with an optimal sampling interval, and we highlight the exact coincidence between the negative autocorrelation inherent in laser chaos and decision-making performance. This study paves the way for a new realm of ultrafast photonics in the age of AI, where the ultrahigh bandwidth of photons can provide new value.

# INTRODUCTION

Physical unique attributes of photons have been utilized in information processing in the literature of optical computing[1]. New photonic processing principles have recently emerged to solve complex time-series prediction problems[2-4], and issues in spatiotemporal dynamics[5] and combinatorial optimization[6], which coincide with the rapid shift to the age of artificial intelligence (AI). These novel approaches exploit the ultrahigh bandwidth attributes of photons and their enabling device technologies[2,3,6]. This paper experimentally demonstrates the usefulness of ultrafast chaotic oscillatory dynamics in semiconductor lasers for reinforcement learning, which is among the most important elements in machine learning.

Reinforcement learning involves adequate decision making in dynamic and uncertain environments[7]. It forms the foundation of a variety of applications, such as information infrastructures[8], online advertisements[9], robotics[10], transportation[11], and Monte Carlo tree search[12], which is used in computer gaming[13]. A fundamental of reinforcement learning is known as the multi-armed bandit problem (MAB), where the goal is to maximize total reward from multiple slot machines, the reward probabilities of which are unknown[7,14,15]. To solve the MAB, one needs to explore better slot machines. However, too much exploration may result in excessive loss, whereas too quick a decision, or insufficient exploration, may lead to neglect of the best machine. There is a trade-off, referred to as the *exploration–exploitation dilemma*[7]. A variety of algorithms for solving the MAB have been proposed in the literature, such as ε-greedy[14], softmax[16], and upper confidence bound[17].

These approaches typically involve probabilistic attributes, especially for exploration purposes. While the implementation and improvements of such algorithms on conventional digital computing are important for various practical applications, understanding their limitations and investigating novel approaches are also important based on perspectives from postsilicon computing. For example, the pseudo-random number generation (RNG) used in conventional algorithmic approaches has severe limitations, such as its data rate, due to the operating frequencies of digital processors (~ GHz range). Moreover, the quality of randomness in RNG has serious limitations[18]. The usefulness of photonic random processes for machine learning is also discussed by utilizing multiple optical scattering[19].

We consider that *directly* utilizing physical irregular processes in nature is an exciting approach with the goal of realizing artificially constructed, physical decision-making machines[20]. Indeed, the intelligence of slime moulds or amoebae, a single-cell natural organism, has been used in solution searches, whereby complex inter-cellular spatiotemporal dynamics play a key role[21]. This stimulated the subsequent discovery of a new principle of decision making strategy called tug of war (TOW), invented by Kim *et al.*[22,23]. The principle of the TOW method originated from the observation of slime moulds: the dynamic expanding and shrinking of their bodies while maintaining a constant intracellular resource volume allows them to collect environmental information, and the conservation of the volume of their bodies entails a nonlocal correlation within the body. The fluctuation, or probabilistic behaviour, in the body of amoebae is important for the exploration of better solutions. The name "TOW" is a metaphor to represent such a *nonlocal* correlation while accommodating *fluctuation*, which enhances decision-making performance[23].

This principle can be adapted to photonic processes. In past research, we experimentally exhibited physical decision making based on near-field-mediated optical excitation transfer at the nanoscale[20,24] and with single photons[25]. These former studies pursued the ultimate physical attributes of photons in terms of diffraction limit-free spatial resolutions and energy efficiency by near-field photons[26,27], and the quantum attributes of single-light quanta[28]. The *nonlocal* aspect of TOW is *directly* physically represented by the wave nature of a single photon or an exciton polariton, whereas fluctuation is also directly represented by their intrinsic probabilistic attributes. However, the fluctuations are limited by practical limitations on the measurements and control systems (second range in the worst case) as well as the single-photon generation rate (kHz range).

The ultrafast, high-bandwidth aspect of photons is another promising physical platform for TOW-based decision making to complement diffraction-limit-free and low-energy near-field photon approaches as well as quantum-level single-photon strategies. As demonstrated below, chaotic oscillatory dynamics of lasers that contains negative autocorrelation experimentally demonstrates 1 GHz decision making. In addition to the resultant *speed merit*, it should be emphasized that the technological maturity of ultrafast photonic devices allows for relatively easy and scalable system implementation through commercially available photonic devices. Furthermore, the applications of the proposed ultrafast photonics-based, and the former near-field/single-photon-based, decision making are *complementary*: the former targets high-end, data centre scenarios by highlighting ultrafast performance, whereas the latter appeals to low-energy, Internet-of-Things-related (IoT)[29], and security[30] applications.

In this study, we demonstrate ultrafast reinforcement learning based on chaotic oscillatory dynamics in semiconductor lasers[31-34] that yields adaptation from zero-prior knowledge at a gigahertz (GHz) range. The randomness is based on complex dynamics in lasers[32-34], and its resulting speed is unachievable in other mechanisms, at least through technologically reasonable means. We experimentally show that ultrafast photonics has significant potential for reinforcement learning. The proposed principles using ultrafast temporal dynamics can be matched to applications including an arbitration of resources at data centres[35] and high-frequency trading[36], where decision making is required at least within milliseconds, and other such high-end utilities. Scientifically, this study paves a way toward the understandings of the physical origin of the enhancement of intelligent abilities (which is reinforcement learning herein) when natural processes (laser chaos herein) are coupled with external systems; this is what we call *natural intelligence*.

Chaotic dynamics in lasers have been examined in the literature[32-34], and their applications have exploited the ultrafast attributes of photonics for secure communication[37-39], random number generation[31,40,41], remote sensing[42], and reservoir computing[2-4]. Reservoir computing is a type of neural network similar to deep learning[13] that has been intensively studied to provide recognition- and prediction-related functionalities. The reinforcement learning described in this study differs completely from reservoir computing from the perspective that neither a virtual network nor machine learning for output weights is required. However, it should be noted that reinforcement learning is important in complementing the capabilities of neural networks, indicating the potential for the *fusion* of photonic reservoir computing with photonic reinforcement learning in future work.

**Principle of reinforcement learning**

For the simplest case that preserves the essence of solving the MAB, we consider a player who selects one of two slot machines, called slot machines 1 and 2 hereafter, with the goal of maximizing reward (known as the two-armed bandit problem). Denoting the reward probabilities of the slot machines by $P_i$ $(i = 1, 2)$, the problem is to select the machine with the highest reward probability. The amount of reward dispensed by each slot machine for a play is assumed to be the same in this study.

The measured chaotic signal $s(t)$ is subjected to the threshold adjuster (TA), according to the TOW principle. The output of the TA is *immediately* the decision concerning the slot machine to choose. If $s(t)$ is equal to or greater than the threshold value $T(t)$, the decision is made to select slot machine 1. Otherwise, the decision is made to select slot machine 2. The reward—the win/lose information of a slot machine play—is fed back to the TA.

The chaotic signal level $s(t)$ is compared with the threshold value $T(t)$ denoted by

$$T(t) = k \times \lfloor TA(t) \rfloor \quad (1)$$

where $TA(t)$ is the threshold adjuster value at cycle $t$, $\lfloor TA(t) \rfloor$ is the nearest integer to $TA(t)$ rounded to zero, and $k$ is a constant determining the range of the resultant $T(t)$. In this study, we assumed that $\lfloor TA(t) \rfloor$ takes the values $-N, \cdots -1, 0, 1, \cdots, N$, where $N$ is a natural number. Hence the number of the thresholds is $2N+1$, referred to as TA's resolution. The range of $T(t)$ is limited between $-kN$ and $kN$ by setting $T(t) = kN$ when $\lfloor TA(t) \rfloor$ is greater than $N$, as well as $T(t) = -kN$ when $\lfloor TA(t) \rfloor$ is smaller than $-N$.

If the selected slot machine yields a reward at cycle $t$ (in other words, wins the slot machine play), the TA value is updated at cycle $t+1$ based on

$$\begin{aligned} TA(t+1) &= -\Delta + \alpha TA(t) \quad \text{if slot machine 1 wins} \\ TA(t+1) &= +\Delta + \alpha TA(t) \quad \text{if slot machine 2 wins} \end{aligned} \quad (2)$$

where $\alpha$ is referred to as the forgetting (memory) parameter[20], and $\Delta$ is the constant increment (in this experiment, $\Delta = 1$ and $\alpha = 0.999$). In this study, the initial TA value was zero. If the selected machine does *not* yield a reward (or loses in the slot machine play), the TA value is updated by

$$\begin{aligned} TA(t+1) &= +\Omega + \alpha TA(t) \quad \text{if slot machine 1 fails} \\ TA(t+1) &= -\Omega + \alpha TA(t) \quad \text{if slot machine 2 fails} \end{aligned} \quad (3)$$

where $\Omega$ is the increment parameter defined below. Intuitively speaking, the TA takes a smaller value if slot machine 1 is considered more likely to win, and a greater value if slot machine 2 is considered more likely to earn the reward. This is as if the TA value is being *pulled* by the two slot machines at both ends, which coincides with the notion of a tug of war.

The fluctuation, necessary for exploration, is realized by associating the TA value with the *threshold* of digitization of the chaotic signal train. If the chaotic signal level $s(t)$ is equal to or greater than the assumed threshold $T(t)$, the decision is immediately made to choose slot machine 1; otherwise, the decision is made to select slot machine 2. Initially, the threshold is zero; hence, the probability of choosing either slot machine 1 or 2 is 0.5. As time elapses, the TA value *shifts* (becomes positive or negative) towards the slot machine with the higher reward probability based on the dynamics shown in Eqs. (2) and (3). We should note that due to the irregular nature of the incoming chaotic signal, the possibility of choosing the *opposite* machine is not zero, which is a critical feature of exploration in reinforcement learning. For example, even when the TA value is sufficiently small (meaning that slot machine 1 seems highly likely to be the better machine), the probability of the decision to choose slot machine 2 is *not* zero.

In TOW-based decision making, the increment parameter $\Omega$ in Eq. (3) is determined based on the history of betting results. Let the number of times when slot machine $i$ is selected in cycle $t$ be $S_i$ and the number of wins in selecting slot machine $i$ be $L_i$. The estimated reward probabilities of slot machines 1 and 2 are given by

$$\hat{P}_1 = \frac{L_1}{S_1}, \ \hat{P}_2 = \frac{L_2}{S_2}. \tag{4}$$

$\Omega$ is then given by

$$\Omega = \frac{\hat{P}_1 + \hat{P}_2}{2 - (\hat{P}_1 + \hat{P}_2)}. \tag{5}$$

The initial $\Omega$ is assumed to be unity, and a constant value is assumed when the denominator of Eq. (5) is zero. The detailed derivation of Eq. (5) is shown in Ref. 23.

**RESULTS**

The architecture of laser chaos-based reinforcement learning is schematically shown in Fig. 1a. A semiconductor laser is coupled with a polarization-maintaining (PM) coupler. The emitted light is incident on a variable fibre reflector, by which a delayed optical feedback is supplied to the laser, leading to laser chaos[40]. The output light at the other end of the PM coupler is detected by a high-speed, AC-coupled photodetector through an optical isolator (ISO) and an attenuator, and is sampled by a high-speed digital oscilloscope at a rate of 100 GSample/s (a 10-ps sampling interval). The detailed specifications of the experimental apparatus are described in the *Methods* section.

Figure 1b shows an example of the chaotic signal train. Figures 1c and 1d show the optical and radio frequency (RF) spectra of laser chaos measured by optical and RF spectrum analysers, respectively. The semiconductor laser was operated at a centre wavelength of 1547.782 nm. The standard bandwidth[31] of the RF spectrum was estimated as 10.8 GHz. Figure 1e summarizes the histogram of the signal levels of the chaotic trains, which spanned from −0.2 to 0.2, with the level zero at its maximum incidence, and slightly skewed between the positive and negative sides. A remark is that the small incidence peak at the lowest measured amplitude is due to our experimental apparatus (not by the laser), and it does not critically effect in the present study. The rounded TA values, $\lfloor TA(t) \rfloor$, are also schematically illustrated at the right-hand side of Fig. 1b and the upper side of Fig. 1e, assuming $N = 10$. This means that $\lfloor TA(t) \rfloor$ ranges from $-10$ to $10$. Signal value $s(t)$ spans between approximately $-0.2$ and $0.2$. The particular example of TA values shown in Figs. 1b and 1e shows the case when the constant $k$ in Eq. (1) is given by 0.02, so that the actual threshold $T(t)$ spans from $-0.2$ to $0.2$. In the experimental demonstration of this study, the TA and the slot machines were emulated in offline processing, whereas online processing was technologically feasible due to the simple procedure of the TA (Remarks are given in the *Discussion* section).

**[EXPERIMENT-1] Adaptation to sudden environmental changes**

We first solved the two-armed bandit problems given by the following two cases, where the reward probabilities $\{P_1, P_2\}$ were given by $\{0.8, 0.2\}$ and $\{0.6, 0.4\}$. We assumed that the sum of the reward probabilities $P_1 + P_2$ was known prior to slot machine plays. With this knowledge, $\Omega$ is unity by Eq. (5).

The slot machine was consecutively played 4,000 times, and this play was repeated 100 times. The red and blue curves in Fig. 2a show the evolution of the *correct decision rate* **(CDR),** defined by the ratio of the number of selections of the machines that yielded the higher reward probability at cycle $t$ in 100 trials, with respect to the probability combination of $\{0.8, 0.2\}$ and $\{0.6, 0.4\}$. The chaotic signal was sampled every 10 ps; hence, the total duration of the 4,000 plays of the slot machine was 40 ns. In order to represent sudden environmental changes (or uncertainty), the reward probability was forcibly interchanged every 10 ns, or every 1,000 plays (For example, $\{P_1, P_2\} = \{0.8, 0.2\}$ was reconfigured to $\{P_1, P_2\} = \{0.2, 0.8\}$). The *resolution* of TA was set to 5 (or $N = 2$).

We observed that the CDR curves quickly approached unity, even after sudden changes in the reward probabilities, showing successful decision making. The adaptation was steeper in the case of a reward probability combination of $\{0.8, 0.2\}$ than that of $\{0.6, 0.4\}$, since the difference in the reward probability was greater in the former case ($0.8 - 0.2 = 0.6$) than the latter ($0.6 - 0.4 = 0.2$). This meant that decision making was easier. The red and blue curves in Fig. 2b represent the evolution of TA values ($TA(t)$) in the case of $\{0.8, 0.2\}$ and $\{0.6, 0.4\}$, respectively, where the TA value became greater than 2 or $-2$; hence, it took long for the TA values to be inverted to the other polarity following environmental change. This is the origin of some *delay* in the CDR in responding to the environmental changes observed in Fig. 2a. One simple method of improving adaptation is to limit the maximum and minimum values of $TA(t)$. The forgetting parameter $\alpha$ is also crucial to improving adaptation speed.

Figures 2c, 2d and 2e characterize decision-making performance dependencies with respect to the configuration of TA. Figure 2c concerns *TA resolution*s demonstrated by the seven curves therein, with corresponding TA resolutions of 5 to 255, where the adaptation was quicker in fewer resolutions than not. Figure 2d considers *TA range* dependencies: while keeping the centre of the TA value at zero and the TA resolution at 5, the three curves in Fig. 2d compare CDRs by a TA range of 0.1, 0.2 and 0.4. We observed that full coverage (0.4) of the chaotic signal yields the best performance. Figure 2e shows the *TA's centre value* dependencies while maintaining TA ranges of 0.4 and a resolution of 5, where we can clearly observe the deterioration of CDRs by the shift in the centre of the TA from zero.

**[EXPERIMENT-2] Adaptation from zero prior knowledge**

Here we considered decision-making problems *without any prior knowledge* of the slot machines. Hence, parameter $\Omega$ needed to be updated. The reward probabilities of the slot machines were set to $\{P_1, P_2\} = \{0.5, 0.1\}$, and 50 consecutive plays were executed. The colour curves in Fig. 3a depict CDRs at six sampling intervals of chaotic signal trains from 10 to 400 ps. Hence, the time needed to complete 50 consecutive slot machine plays differed among these; it ranges from $10 \text{ ps} \times 50 = 500 \text{ ps}$ to $400 \text{ ps} \times 50 = 20 \text{ ns}$. Moreover, the black curve shows the CDR obtained by uniformly distributed pseudo-random numbers generated by the Mersenne Twister for the random signal source, instead of experimentally observed chaotic signals. We observed that CDRs based on chaotic signals exhibited more rapid adaptation than the uniformly distributed pseudo-random numbers.

The most prompt adaptation, or the optimal performance, was obtained at a particular sampling interval. Figure 3b compares CDRs at the cycle $t = 3$; the sampling interval of 50 ps yielded the best performance, indicating that the original chaotic dynamics of the laser could *physically* be optimized such that the most prompt decision-making is realized. Indeed, the *autocorrelation* of the laser chaos signal trains was evaluated as shown in Fig. 3c, its negative maximum value is taken when the time lag is given by 5 or −5, corresponding *exactly* to the sampling interval of 50 ps ($5 \times 10$ ps). In other words, the negative correlation of chaotic dynamics enhanced the exploration ability for decision making. Furthermore, this finding suggests that optimal performance is obtained at the maximum sampling rate (or data rate) by physically tuning the dynamics of the original laser chaos, which will be an important and exciting topic for future investigation. The adaptation speed of decision making was estimated as 1 GHz in this optimal case, where CDR was larger than 0.95 at 20 cycles with a 50-ps sampling interval (20 GSamples/s) in Fig. 3a (1 GHz = (50 ps × 20 cycles)$^{-1}$).

Furthermore, we characterized CDRs with normally distributed random numbers (referred to as RANDN) in order to ensure that the statistical incidence patterns of the laser chaos, which were similar to a normal distribution shown in Fig. 1e, were *not* the origins of the fast adaptation of the decision making. By keeping the mean value of RANDN at zero, the standard deviation ($\sigma$) was configured as 0.2, 0.1 and 0.01. As shown in Fig. 4, the CDRs at cycle $t = 3$ by RANDN were inferior to chaotic signals and the uniformly distributed random numbers (denoted by RAND). Moreover, the CDR was evaluated by *surrogating* the chaotic signal time series sampled at 50 ps intervals, which resulted in *poorer* performance than the original, as shown in Fig. 4. These evaluations support the claim that laser chaos is beneficial to the performance of reinforcement learning, in addition to its ultrafast data rate for random signals.

**DISCUSSION**

The performance enhancement of decision making by chaotic laser dynamics is demonstrated and the impacts of negative autocorrelation is clearly suggested. Further understanding between chaotic oscillatory dynamics and decision making is a part of important future research. The first regards to physical insights. Toomey *et al.* recently showed that the complexity of laser chaos varies within the coherence collapse region in the given system[43]. The level of the optical feedback, injection current of the laser becomes an important parameter in determining the complexity of chaos, which is the entrance to thorough insights. We also consider the use of the bandwidth enhancement technique[31] with optically injected lasers to improve the adaptation speed of decision making over tens of GHz. Meanwhile, besides negative autocorrelation inherent in laser chaos, other perspectives could address the underlying mechanism such as diffusivity,[44] Hurst exponents,[45] etc.

In the experimental demonstration, the *nonlocal* aspect of the TOW principle was *not directly,* physically relevant to the chaotic oscillatory dynamics of the lasers. By *combining* the chaotic dynamics with the threshold adjustor, the nonlocal and fluctuation properties of the TOW principle emerged, which have *not* been completely realized in the literature nor in our past experimental studies[24,25]. Such a hybrid realization of nonlocality in TOW leads to higher likelihood of technological implementability, and better scalability and extension to higher-grade problems. Online post-processing can become feasible through electric circuitry, as already demonstrated for a random-bit generator[41,46] since the post-processing is very simple as described in Eqs. (1) to (5). For scalability, a variety of approaches can be considered, such as time-domain multiplication, which exploits the ultrafast attributes of chaotic lasers, and is a frequently used strategy in ultrafast photonic systems. Introducing multi-threshold values[31,41] is another simple extension of our proposed scheme.

*Whole-photonic realization* is an interesting issue to explore, and has already been implied by the analysis where the time-domain correlation of laser chaos strongly influences decision-making performances. The mode dynamics of multi-mode lasers[47] are very promising for the implementation of nonlocal properties of fully photonic systems required for decision making. Synchronization and its clustering properties in coupled laser networks[48,49] are also interesting approaches to physically realizing the nonlocality of the TOW. These systems can automatically tune the optimal settings for decision making (e.g., negative autocorrelation properties), which can lead to autonomous photonic intelligence.

We also make note of the *extension* of the principle demonstrated in this paper to *higher-grade* machine learning problems. The *competitive MAB*[50] with multiple players is an exciting topic for photonic intelligence research as it involves the so-called Nash equilibrium, and is the foundation of such important applications as resource allocation and social optimization. Investigating the possibilities of extending the present method and utilizing ultrafast laser dynamics for competitive MAB is highly interesting.

**Conclusion**

We experimentally established that laser chaos provides ultrafast reinforcement learning and decision making. The adaptation speed of decision making reached 1 GHz in the optimal case with the sampling rate of 20 GSample/s (50-ps decision-making intervals) using the ultrafast dynamics inherent in laser chaos. The maximum adaptation performance coincided with the negative maximum of the autocorrelation of the original time-domain laser chaos sequences, demonstrating the strong impact of chaotic lasers on decision making. The origin of the performance was also validated by comparing with uniformly and normally distributed pseudo-random numbers as well as surrogated arrangements of original chaotic signal trains. This study is the first demonstration of ultrafast photonic reinforcement learning or decision making, to the best of our knowledge, and paves the way for research on photonic intelligence and new applications of chaotic lasers in the realm of artificial intelligence.

**METHODS**

**Optical system**

The laser used in the experiment was a distributed-feedback (DFB) semiconductor laser mounted on a butterfly package with optical fibre pigtails (NTT Electronics, KELD1C5GAAA). The injection current of the semiconductor laser was set to 58.5 mA (5.85 $I_{th}$), where the lasing threshold $I_{th}$ was 10.0 mA. The relaxation oscillation frequency of the laser was 6.5 GHz. The temperature of the semiconductor laser was set to 294.86 K. The laser output power was 13.2 mW. The laser was connected to a variable fibre reflector, which reflected a fraction of light back into the laser, inducing high-frequency chaotic oscillations of optical intensity[32-34]. 1.9 % of the laser output power was fed back to the laser cavity from the reflector. The fibre length between the laser and the reflector was 4.55 m, corresponding to the feedback delay time (round trip) of 43.8 ns. Polarization-maintaining fibres were used for all optical fibre components. The optical output was converted to an electronic signal by a photodetector (New Focus, 1474-A, 35 GHz bandwidth) and sampled by a digital oscilloscope (Tektronics, DPO73304D, 33 GHz bandwidth, 100 GSample/s, eight-bit vertical resolution). The RF spectrum of the laser was measured by an RF spectrum analyser (Agilent, N9010A-544, 44 GHz bandwidth). The optical wavelength of the laser was measured by an optical spectrum analyser (Yokogawa, AQ6370C-20).

**Data analysis**

**[EXPERIMENT-1]** A chaotically oscillated signal train was sampled at a rate of 100 GSample/s by 9,999,994 points, which lasted approximately 10 μs. As described in the main text, 4,000 consecutive plays were repeated 100 times; hence, the total number of slot machine plays was 400,000. With a 10-ps interval sampling, the initial 400,000 points of the chaotic signal were used for the decision-making experiments. The post-processing required for 400,000 iterations of slot machine plays was approximately 1.2 s (or 3.0 μs/decision), on a normal-grade personal computer (Panasonic, CF-SX3, 16 GB RAM, Windows 7, MATLAB R2011b).

**[EXPERIMENT-2]**

**(1) Sampling methods:** A chaotic signal train was sampled at 10-ps intervals with 9,999,994 sampling points. Such a train was measured 120 times. Each chaotic signal train was referred to as $chaos_i$, and there were 120 kinds of such trains: $i = 1, \cdots, 120$. In demonstrating $10 \times M$ ps sampling intervals, where $M$ was a natural number ranging from 1 to 40 (namely, the sampling intervals were 10 ps, 20 ps, $\cdots$, and 400 ps), we chose one of every $M$ samples from the original sequence.

**(2) Evaluation of CDR regarding a specific chaos sequence:** For every chaotic signal train $chaos_i$, 50 consecutive plays were repeated 200 times. Consequently, 10,000 points were used from $chaos_i$. Such evaluations were repeated 100 times. Hence, 1,000,000 slot machine plays were conducted in total. These CDRs were calculated for all signal trains $chaos_i$ ($i = 1, \cdots, 120$).

**(3) Evaluation of CDR of all chaotic sequences:** We evaluated the average CDR of all chaotic signal trains ($i = 1, \cdots, 120$) derived in **(2)** above, which were the results discussed in the main text.

**(4) Autocorrelation of chaotic signals:** The autocorrelation was computed based on all 9,999,994 sampling points of $chaos_i$, and was evaluated for all $chaos_i$ ($i = 1, \cdots, 120$). The autocorrelation demonstrated in Fig. 3c was evaluated as the average of these 120 kinds of autocorrelations.

**(5) Surrogate methods:** The surrogate time series of original chaotic sequences were generated by the `randperm` function in MATLAB which is based on the sorting of pseudorandom numbers generated by Mersenne Twister.

**Data availability.** The data sets generated during the current study are available from the corresponding author on reasonable request.

**Acknowledgements**

This work was supported in part by the Core-to-Core Program, A. Advanced Research Networks from the Japan Society for the Promotion of Science and Grants-in-Aid for Scientific Research from Japan Society for the Promotion of Science.

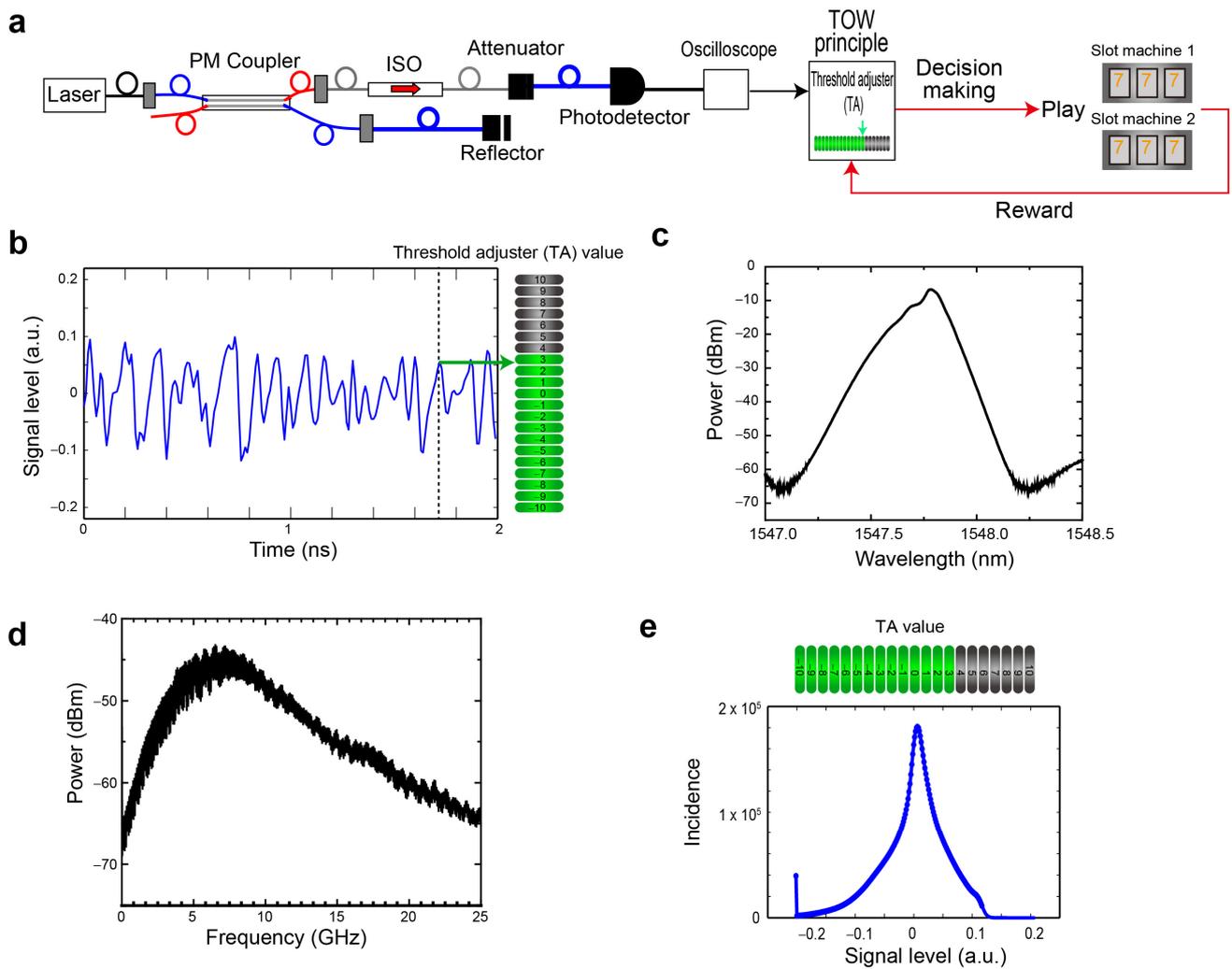

**Figure 1 | Architecture of photonic reinforcement learning based on laser chaos. (a)** Architecture and experimental configuration of laser chaos-based reinforcement learning. Ultrafast chaotic optical signal is subjected to the tug-of-war (TOW) principle that determines the selection of slot machines. ISO: optical isolator. **(b)** An example of chaotic signal trains sampled at 100 GSample/s. The signal level is subjected to a threshold adjustment (TA) for decision making. **(c)** Optical spectrum and **(d)** RF spectrum of the laser chaos used in the experiment. **(e)** Incidence statistics (histogram) of the signal level of the laser chaos signal.

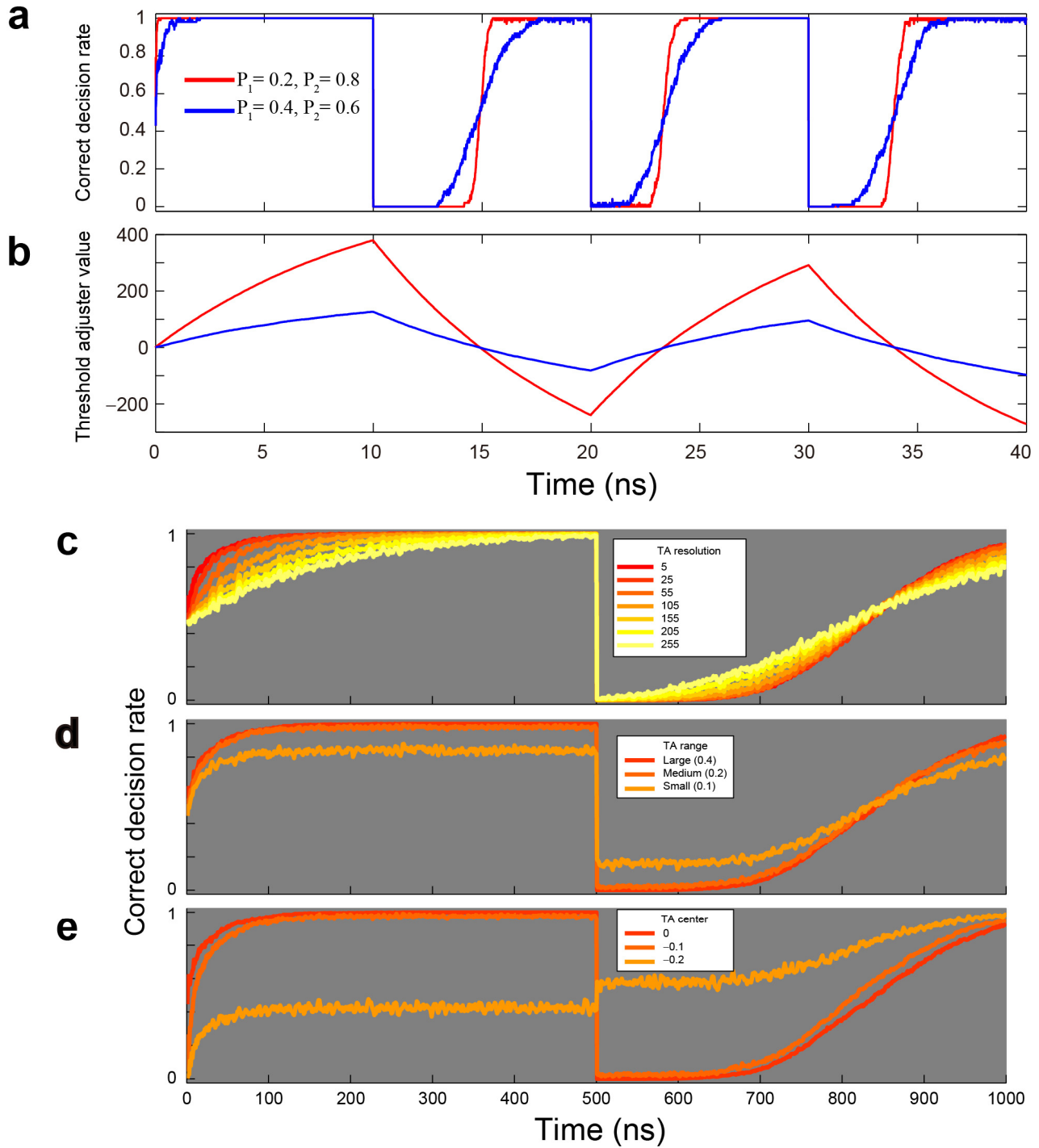

**Figure 2. Reinforcement learning in dynamically changing environments. (a)** Evolution of the correct decision rate (CDR) when the reward probabilities of the two slot machines are {0.8, 0.2} and {0.6, 0.4}. The knowledge that the sum of the reward probability, which is unity in these cases, is supposed to be given. The reward probability is intentionally swapped every 10 ns in order to represent sudden environmental changes or uncertainty. Rapid and adequate adaptation is observed in both cases. **(b)** Evolution of the threshold adjuster (TA) value underlying correct decision making. **(c-e)** CDR performance dependency on the setting of TA. **(c)** TA resolution dependency. **(d)** TA range dependency. **(e)** TA centre value dependency.

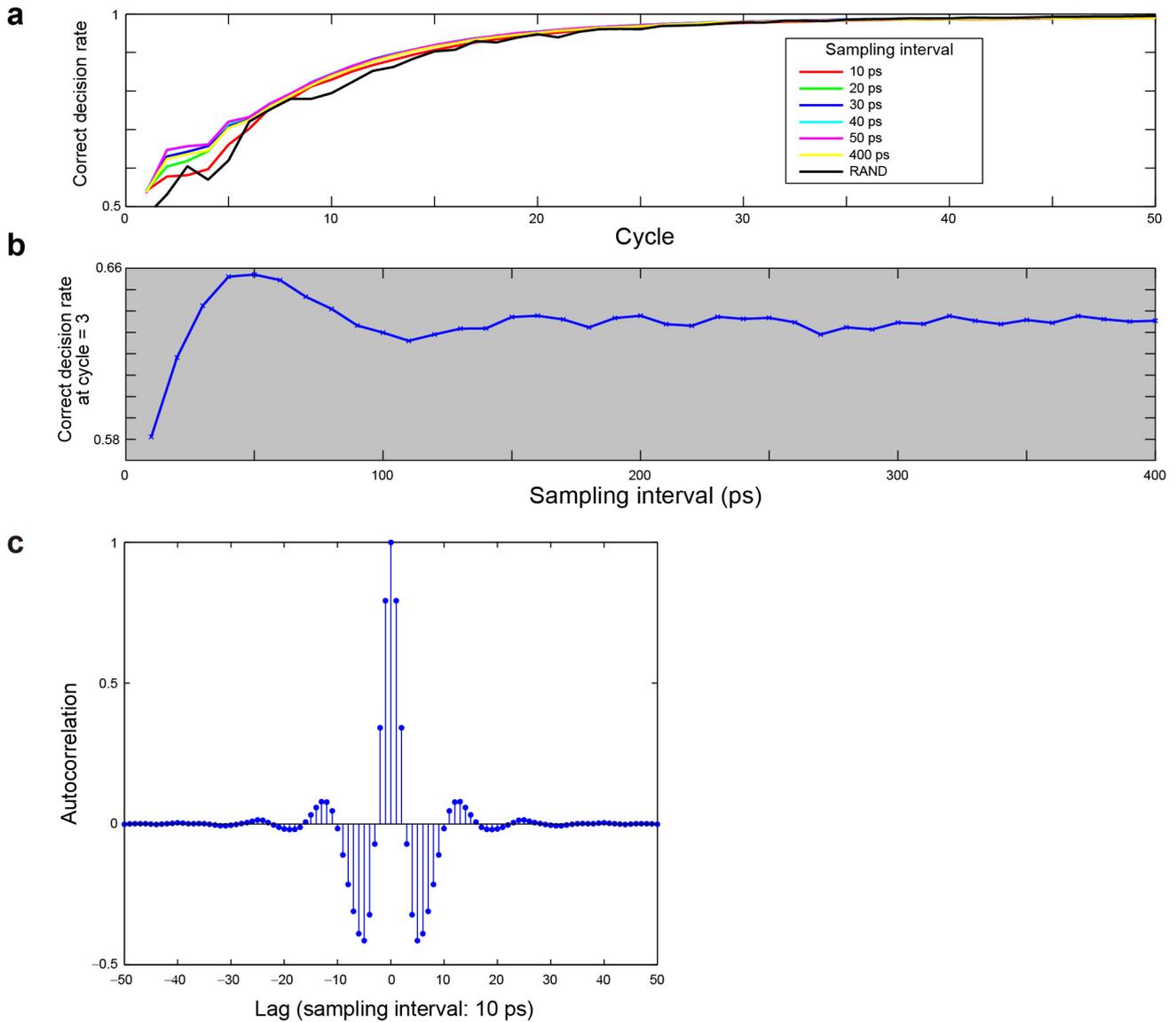

**Figure 3 | Reinforcement learning from *zero* prior knowledge.** **(a)** Evolution of CDR with different sampling intervals of the laser chaos signal (10, 20, 30, 40, 50, and 400 ps) and uniformly distributed pseudo-random numbers. CDR exhibits prompt adaptation when the sampling interval is 50 ps. **(b)** CDR is evaluated as a function of the sampling interval from 10 ps to 400 ps, where the maximum performance is obtained at 50 ps. **(c)** Autocorrelation of the laser chaos signals exhibits its negative maximum when the time lag is 5 or −5, which exactly coincides with the fact that the optimal adaptation is realized at 50 ps (10 ps × 5) sampling intervals.

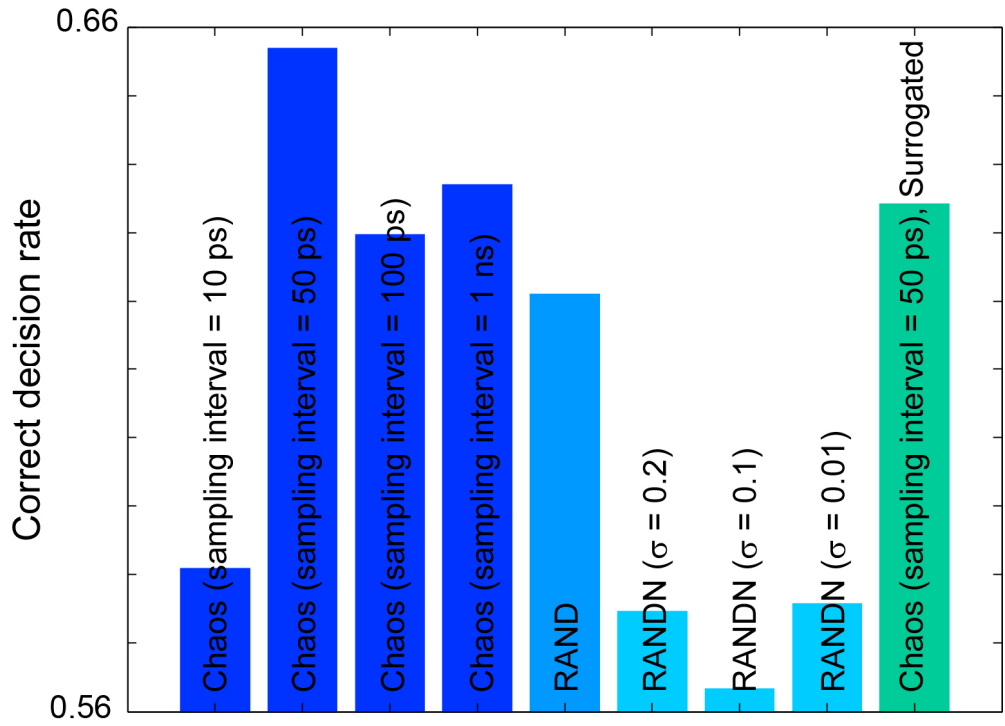

**Figure 4 | Comparison of learning performance in laser chaos, uniformly and normally distributed pseudo-random numbers, and surrogate laser chaos signals.** The laser chaos sampled at 50 ps interval exhibits the best performance compared with other cases, indicating that the dynamics of laser chaos affects reinforcement learning ability.